\documentclass[12pt]{iopart}
\usepackage{iopams}  
\usepackage{bm}
\usepackage{bbm}
\usepackage{graphics}

\begin{document}

\title[\small{Polarization dynamics in twisted fiber amplifiers: a non-Hermitian nonlinear dimer model}]{Polarization dynamics in twisted fiber amplifiers: \\ a non-Hermitian nonlinear dimer model}

\author{J. D. Huerta Morales$^{1}$, B. M. Rodr\'iguez-Lara$^{1,2,*}$ and B. A. Malomed$^{3,4}$}

\address{$^{1}$ \quad Instituto Nacional de Astrof\'{\i}sica, \'Optica y Electr\'onica, Calle Luis Enrique Erro No. 1, Sta. Ma. Tonantzintla, Pue. CP 72840, M\'exico; jd\_huerta@inaoep.mx\\
	$^{2}$ \quad Tecnol\'ogico de Monterrey, Escuela de Ingenier\'ia y Ciencias, Ave. Eugenio Garza Sada 2501, Monterrey, N.L. 64849, M\'exico \\
    $^{3}$ \quad Department of Physical Electronics, School of Electrical Engineering, Faculty of Engineering and Tel Aviv University Center for Light-Matter Interaction, Tel Aviv 69978, Israel\\
    $^{4}$ \quad ITMO University, St. Petersburg 197101, Russia}
\ead{bmlara@itesm.mx}
\vspace{10pt}

\begin{abstract}
We study continuous-wave light propagation through a twisted birefringent single-mode fiber amplifier with saturable nonlinearity. 
The corresponding coupled-mode system is isomorphic to a non-Hermitian nonlinear dimer, and gives rise to analytic polarization-mode dynamics. It provides an optical simulation of the semi-classical non-Hermitian Bose-Hubbard model, and suggests its use for the design of polarization circulators and filters, as well as sources of polarized light. \end{abstract}

%
%
%
%
%

\section{Introduction}

Single-mode fiber amplifiers exhibit a plethora of nonlinear and dispersive effects \cite{TerMikirtychev2014}, in addition to the amplification proper, that makes them a wide-range platform for device development \cite{Okhotnikov2012}.
Here, we show new possibilities for simulating \textit{exotic physics} and application-oriented device design, using continuous-wave light propagation through a twisted birefringent single-mode fiber amplifier.
The natural and mechanically induced birefringence of the fiber provide for polarization mode-coupling, which has been used to design polarization rotators, polarization-mode interchangers, and delay equalizers \cite{Ulrich1979p2241,Monerie1980p449,Fang1987p978,Almanee2016p4927}.
Related to this, we assume polarization instability that might be caused by rare-earth doping and Kerr nonlinearity \cite{Zeghlache1995p4229,Lelarge2007p111,Tentori2013p31725}.

We use mode-coupling theory to describe the propagation of orthogonally polarized spatial modes through such a fiber:
\begin{eqnarray} \label{815eq1}
-i{{\partial }_{z}}{{\mathcal{E} }_{1}}&=&{{\alpha }_{1}}{{\mathcal{E}}_{1}}+g{{e}^{iqz}}{{\mathcal{E}}_{2}}+\left( {{a}_{1}}{{\left| {{\mathcal{E}}_{1}} \right|}^{2}}+{{b}_{1}}{{\left| {{\mathcal{E}}_{2}} \right|}^{2}} \right){{\mathcal{E}}_{1}}, \nonumber  \\ 
-i{{\partial }_{z}}{{\mathcal{E}}_{2}}&=&{{\alpha }_{2}}{{\mathcal{E}}_{2}}+g{{e}^{-iqz}}{{\mathcal{E}}_{1}}+\left( {{b}_{2}}{{\left| {{\mathcal{E}}_{1}} \right|}^{2}}+{{a}_{2}}{{\left| {{\mathcal{E}}_{2}} \right|}^{2}} \right){{\mathcal{E}}_{2}},
\end{eqnarray} 
where the complex field amplitudes $\mathcal{E}_{j} \equiv \mathcal{E}_{j}(z)$ represent the polarized field modes, the real and imaginary parts of the complex parameters $\alpha_{j}$ being related to the effective refractive indices \cite{Feldman1993p1191} and linear gain for each mode, respectively.
The real amplitude, $g$, and phase, $q$, of the effective inter-modal coupling, the parameters related to the Kerr nonlinearities, $a_{j}$, and nonlinear couplings, $b_{j}$,  depend on the natural birefringence, the nonlinear ellipse rotation, Kerr response, the mechanical twist of the fiber, and the overlap of the spatial modes \cite{Ulrich1979p2241,Feldman1993p1191}. 
Evanescently coupled waveguides with linear gain or loss give rise to similar propagation models; for example, neglecting the parameter related to the mechanical twist, $q=0$, reduces the model to that of asymmetric active couplers  \cite{Kominis2016p33699,Kominis2017p063832}. Furthermore, null nonlinear response, $a_{j}=b_{j}=0$, reduces the model to the standard $\mathcal{PT}-$symmetric dimer \cite{RodriguezLara2015p5682,HuertaMorales2016p83}, and negligible nonlinear couplings, $b_{j}=0$, to a special type of non-Hermitian nonlinear dimer \cite{Xu2015p055101}.
\section{Polarization dynamics}
In the following, we add direct saturation to the nonlinearity \cite{Tutt1993p299}:
\begin{eqnarray}
-i{{\partial }_{z}}{{\mathcal{E}}_{1}}&=&{{\alpha }_{1}}{{\mathcal{E}}_{1}}+g{{e}^{iqz}}{{\mathcal{E}}_{2}}+\frac{{{a}_{1}}{{\left| {{\mathcal{E}}_{1}} \right|}^{2}}+{{b}_{1}}{{\left| {{\mathcal{E}}_{2}} \right|}^{2}}}{{{\left| {{\mathcal{E}}_{1}} \right|}^{2}}+{{\left| {{\mathcal{E}}_{2}} \right|}^{2}}}{{\mathcal{E}}_{1}}, \nonumber \\ 
-i{{\partial }_{z}}{{\mathcal{E}}_{2}}&=&{{\alpha }_{2}}{{\mathcal{E}}_{2}}+g{{e}^{-iqz}}{{\mathcal{E}}_{1}}+\frac{{{b}_{2}}{{\left| {{\mathcal{E}}_{1}} \right|}^{2}}+{{a}_{2}}{{\left| {{\mathcal{E}}_{2}} \right|}^{2}}}{{{\left| {{\mathcal{E}}_{1}} \right|}^{2}}+{{\left| {{\mathcal{E}}_{2}} \right|}^{2}}}{{\mathcal{E}}_{2}}.
\end{eqnarray}
This modification allows us to elaborate an analytically tractable system that covers both fiber and coupled-waveguides models. We address the dynamics of a Stokes-like vector to demonstrate that our model is tantamount to a classical non-Hermitian Bose-Hubbard model. 
It describes in detail the common scenario of a twisted weakly birefringent fiber, where the dynamics feature a nonlinear bifurcation leading to localized oscillations on the Poincar\'e sphere (PS) \cite{Agrawal2013,Daino1986p42}.
We also consider an instance where the linear gain induces dynamic regimes, characterized by unstable and stable spirals, that can be used to design polarization filters with amplification. 
Finally, we discuss how our analytic model might help design polarization circulators and filters, as well as amplifying sources of polarized light.

We introduce renormalized fields as suggested by the saturable nonlinearity,
\begin{eqnarray}
\mathcal{A}_{j}(z) = \frac{  e^{- i l_{0} \int_{0}^{z} \frac{ \left| {{\mathcal{E}}_{1}}(t) \right|^{2} -  \left| {{\mathcal{E}}_{2}}(t) \right|^{2} } { \left|{\mathcal{E}_{1}}(t) \right|^{2} +  \left| {{\mathcal{E}}_{2}}(t) \right|^{2}}  dt ~  -i \left[ k_{0}  + \alpha_{0} - \frac{(-1)^{j}}{2} q \right] z }} {\sqrt{\left| e^{-i  \alpha_{0} z} {{\mathcal{E}}_{1}}(z) \right|^{2} +  \left|  e^{-i  \alpha_{0} z} {{\mathcal{E}}_{2}}(z) \right|^{2}}}  \mathcal{E}_{j}(z),
\end{eqnarray}
and obtain a modified set of effective coupled-mode equations describing a non-Hermitian nonlinear dimer,
\begin{eqnarray}
- i \partial_{z} \mathcal{A}_{1} &=& g \mathcal{A}_{2} + \left[ n_{r}  + k \left| \mathcal{A}_{1} \right|^2 + \left( 2 i n_{i}  - k \right) \left| \mathcal{A}_{2} \right|^2  \right] \mathcal{A}_{1}, \nonumber \\
- i \partial_{z} \mathcal{A}_{2} &=& g \mathcal{A}_{1} - \left[ n_{r} + \left(2 i n_{i} + k \right) \left| \mathcal{A}_{1} \right|^2 - k \left| \mathcal{A}_{2} \right|^2  \right] \mathcal{A}_{2}, \label{eq:RenormalizedFiberAmplifier}
\end{eqnarray}
with four effective real parameters $g$, $n_{r} =  (\alpha_{1r} - \alpha_{2r})/2 + l - \frac{q}{2}$ with $l=(a_{1}-a_{2}+b_{1}-b_{2})/4$, $n_{i} = (\alpha_{1i} - \alpha_{2i})/2$, and $k=(a_{1}+a_{2}-b_{1}-b_{2})/4$.
A scaled propagation distance, $g z$, can always be used to reduce the effective parameter number to three.
We use the notation $\alpha_{jr}$ and $\alpha_{ji}$ for the real and imaginary parts of $\alpha_{j}$, in that order.
Additionally, we define the auxiliary complex, $\alpha_{0} = (\alpha_{1} + \alpha_{2})/2$, and real, $k_{0}= (a_{1}+a_{2}+b_{1}+b_{2})/4$, $l_{0}= (a_{1}-a_{2}-b_{1}+b_{2})/4$, parameters.
Our model is isomorphic to one describing twisted birefringent fibers under the effect of an external magnetic field \cite{Ulrich1979p2241} for parameter settings $n_{i}=a_{j} = b_{j}=0$, and to the model describing nonlinear birefringent fibers \cite{Feldman1993p1191,Kuzin2001p919} for parameters $n_{i}=0$, $b_{1}=b_{2}= 2 a_{1}$, and $a_{2}= a_{1}$.

To study our model dynamics, we define the real Stokes-like vector:
\begin{eqnarray}
S_{x} &=&  \mathcal{A}_{1} \mathcal{A}_{2}^{\ast}  + \mathcal{A}_{1}^{\ast} \mathcal{A}_{2} , \nonumber \\
S_{y} &=&  i \left( \mathcal{A}_{1} \mathcal{A}_{2}^{\ast} - \mathcal{A}_{1}^{\ast} \mathcal{A}_{2}  \right), \nonumber \\
S_{z} &=&  \left| \mathcal{A}_{1} \right|^2 - \left| \mathcal{A}_{2} \right|^2.
\end{eqnarray}
The evolution of this unit norm vector, $S_{x}^2 + S_{y}^2 + S_{z}^2 =  \left| \mathcal{A}_{1} \right|^2 + \left| \mathcal{A}_{2} \right|^2 = 1$, is governed by equations obtained from Eq. (\ref{eq:RenormalizedFiberAmplifier}):
\begin{eqnarray}
\partial_{z}S_{x} &=& 2 n_{r} S_{y} + 2 \left( n_{i} S_{x} + k S_{y} \right) S_{z}  , \nonumber \\
\partial_{z}S_{y} &=& -2 \left( n_{r} + k S_{z} \right) S_{x} + 2 \left( g + n_{i} S_{y} \right) S_{z}, \nonumber \\
\partial_{z}S_{z} &=&  -2 g S_{y} - 2 n_{i} \left(1 - S_{z}^2 \right). 
\end{eqnarray}
These equations are isomorphic to those of the semi-classical non-Hermitian Bose-Hubbard dimer \cite{Graefe2010p013629}.
Thus, our fiber amplifier may be considered as an optical simulator of the latter condensed matter model, where the Stokes-like vector components represent the semi-classical expectation value of the angular momentum components.

The fixed points of the dynamics, $\mathbf{S}_{j} = (S_{j,x},S_{j,y},S_{j,z})$, determine static equilibria in the model. To this end, we calculate the Jacobian,
\begin{eqnarray}
J = \left( \begin{array}{ccc} 
\partial_{\mathcal{S}_{x}}\partial_{z}S_{x} & 
\partial_{\mathcal{S}_{y}}\partial_{z}S_{x} &
\partial_{\mathcal{S}_{z}}\partial_{\zeta}S_{x} \\
\partial_{\mathcal{S}_{x}}\partial_{z}S_{y} & 
\partial_{\mathcal{S}_{y}}\partial_{z}S_{y} &
\partial_{\mathcal{S}_{z}}\partial_{z}S_{y} \\
\partial_{\mathcal{S}_{x}}\partial_{z}S_{z} & 
\partial_{\mathcal{S}_{y}}\partial_{z}S_{z} & 
\partial_{\mathcal{S}_{z}}\partial_{z}S_{z} 
\end{array}\right),
\end{eqnarray}
at each fixed point,  $J(\mathbf{\mathbf{S}}_{j})$, find its eigenvalues, $\lambda_{j}=(\lambda_{j,1},\lambda_{j,2},\lambda_{j,3})$, and classify them according to the standard dynamical systems theory \cite{Sprott2003}.
This can be done analytically but the results become too cumbersome for the general case.

\section{Examples}
\subsection{Twisted Weakly Birefringent Fiber}

Let us consider the propagation of mutually orthogonal circular polarization modes through a twisted low-birefringent nonlinear fiber without gain as a tractable example \cite{Agrawal2013, Daino1986p42}.
Here, the modal field amplitudes $\mathcal{E}_{1}$ and $\mathcal{E}_{2}$ represent the amplitudes of right- and left-handed circular polarizations. 
The model parameters fulfill $a_{j} = a$, $b_{j}= 2 a$ and $\alpha_{j}$ are real numbers; thus, the effective parameters become $k = -a/2 $ and $n_{r} = \left( \alpha_{1}- \alpha_{2} - q \right)/2$.
In order to provide analytic dynamics, we select the mechanical twist parameter as $q = \alpha_{1} - \alpha_{2}$ such that $n_{r}=0$.
In this case, the dynamics reduce to
\begin{eqnarray}
\partial_{z}S_{x} &=&  - a S_{y} S_{z}  ,  \nonumber \\
\partial_{z}S_{y} &=&    \left( 2 g  + a S_{x}   \right) S_{z}, \nonumber \\
\partial_{z}S_{z} &=& -2 g S_{y},  \label{eq:WeakFiber}
\end{eqnarray}
and gives rise to two fixed points at all parameter values and up to four in parameter regions where the effective coupling and nonlinearity fulfill $d = 2 g / a \le 1$:
\begin{eqnarray}
\mathcal{S}_{1,2} = \left(\pm1,0,0\right), \quad
\mathcal{S}_{3,4} = \left(-d,0,\pm \sqrt{1-d^2} \right). \label{eq:FPWeakFiber}
\end{eqnarray}
The Jacobian eigenvalues evaluated at these fixed points, 
\begin{eqnarray}
\lambda_{1}&=&\left( -\sqrt{- 2 g \left(a + 2 g \right)},0, \sqrt{ -2 g \left(a + 2 g\right)}\right), \nonumber \\
\lambda_{2}&=&\left( -\sqrt{ 2 g \left(a - 2 g \right)}, 0, \sqrt{ 2 g \left( a - 2 g \right)}\right), \nonumber \\
\lambda_{3,4}&=&\left(-\sqrt{4 g^2 -  a^2},0,  \sqrt{4 g^2 -  a^2}\right),
\end{eqnarray}
provide the classification presented in Table \ref{tab:Tab1}. \\
\vspace{-2em}
\begin{table}[!htbp]
	\centering
	\caption{\bf Fixed Point Classification for the System Equivalent to a Weakly Birefringent Fiber Represented by Eq.(\ref{eq:WeakFiber})}
	\begin{tabular}{cccc}
		\hline
		& RI & RII & RIII  \\
		& $ 2g  >  a $ & $  2 g  =  a $ & $  2g  < a $  \\
		\hline
		$\mathcal{S}_{1}$ & Center  & Center 		& Center  \\
		$\mathcal{S}_{2}$ & Center  & Bifurcation  	& Saddle \\
		$\mathcal{S}_{3}$ &   		&   	 		& Center \\
		$\mathcal{S}_{4}$ &   		&  		 		& Center \\
		\hline
	\end{tabular}
	\label{tab:Tab1}
\end{table}

\begin{figure}[htbp]
	\centering
	\fbox{\includegraphics{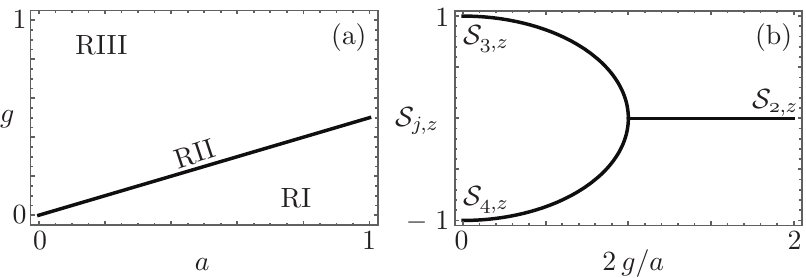}}
	\caption{ (a) Parameter regions defined by the fixed points of the model corresponding to a twisted  weakly birefringent fiber, and (b) the bifurcation of the Rabi oscillation center, $\mathcal{S}_{2}$, into two Josephson oscillations centers, $\mathcal{S}_{3}$ and $\mathcal{S}_{4}$.} 
	\label{fig:Fig1}
\end{figure}

First, we define regime RI where the parameters fulfill $ 2 g  > a $. 
Here, only two fixed points, $\mathcal{S}_{1}$ and $\mathcal{S}_{2}$, exist, of the center type.
In the second (codimension-one) parameter regime, RII, the model parameters satisfy $ 2 g  = a$ and the first fixed point, $\mathcal{S}_{1}$, keeps being a center, while the second, $\mathcal{S}_{2}$, undergoes a bifurcation which transforms it into a saddle point.
The third parameter regime, RIII with  $ 2 g  <  a$, is a generic one, similar to RI.
The first fixed point, $\mathcal{S}_{1}$, remains a center, the second, $\mathcal{S}_{2}$, becomes a saddle point, and there appear two additional fixed points of the center type, one located on the right-handed elliptically polarized hemisphere, $\mathcal{S}_{3}$, and the other, $\mathcal{S}_{4}$, its left-handed counterpart.
Figure \ref{fig:Fig1}(a) shows these three regions in parameter space, while the bifurcation of the center $\mathcal{S}_{2}$ into centers $\mathcal{S}_{3}$ and $\mathcal{S}_{4}$ are presented in Fig. \ref{fig:Fig1}(b). 

\begin{figure}[htbp]
	\centering
	\fbox{\includegraphics{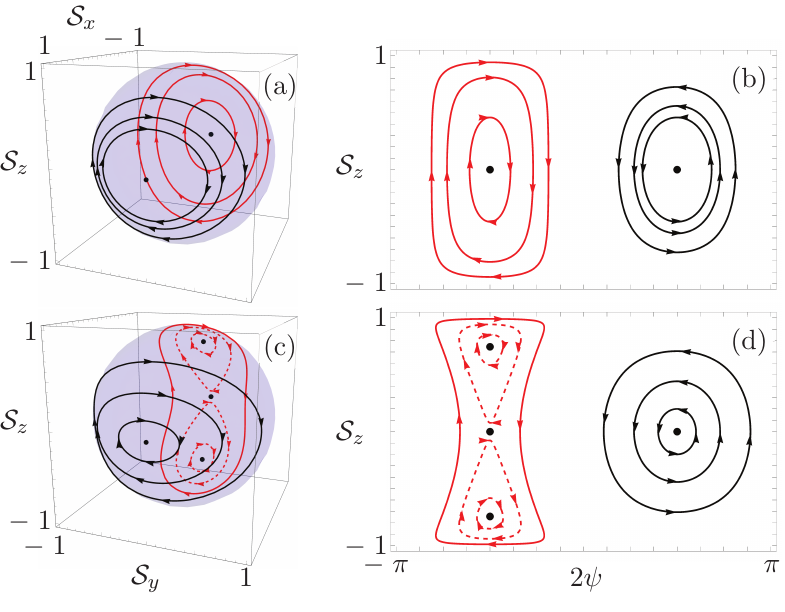}}
	\caption{(Color online) Polarization trajectories on (a),(c) the Poincar\'e sphere and (b),(d) its cylindrical projection, for a weakly birefringent twisted nonlinear fiber in (a) and (b) RI for  $a=g$, and (c) and (d) RIII for $a=3g$. Rabi and Josephson oscillations are shown, respectively, by solid and dashed lines.} 
	\label{fig:Fig2}
\end{figure}

In any parameter region, the fiber acts as a polarization circulator.
For initial polarization ellipse angles, $2 \psi = \arctan \mathcal{S}_{y} / \mathcal{S}_{x}$, in the range $0 < 2 \psi < \pi/2$, the dynamics corresponds to Rabi oscillations around the fixed point $\mathcal{S}_{1}$, see solid black lines in Figs. \ref{fig:Fig2}(a)-\ref{fig:Fig2}(d).
Initial polarizations with angle values in the range $-\pi/2 < 2 \psi < 0$ give rise to Rabi oscillations around the fixed point $\mathcal{S}_{2}$ in RI and RII, see solid red lines, and localized Josephson oscillations around the fixed points $\mathcal{S}_{3}$ and $\mathcal{S}_{4}$ in region RIII for initial right- and left-handed elliptic polarizations, see dashed red lines in Figs. \ref{fig:Fig2}(c) and \ref{fig:Fig2}(d).
In addition to these hemisphere-localized oscillations, the saddle point in region RIII also signals the existence of Rabi oscillations around it, as shown by the solid red lines in Figs. \ref{fig:Fig2}(c) and \ref{fig:Fig2}(d).

\subsection{Twisted Weakly Birefringent Fiber Amplifier}
\begin{figure}[htbp]
	\centering
	\fbox{\includegraphics{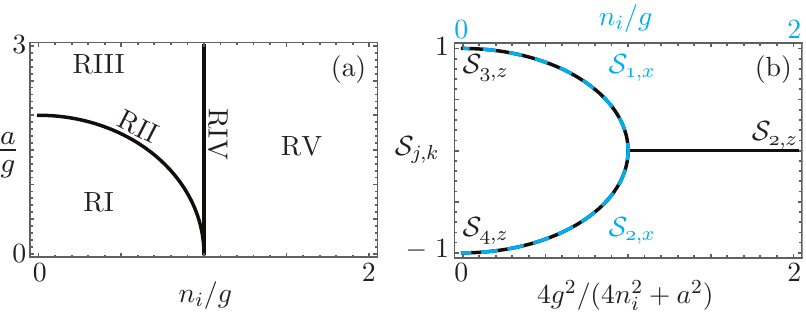}}
	\caption{(Color online) (a) Parameter regions which feature fixed points of different types; see text for details. (b) Some Stokes vector components for fixed points of the unstable and stable spiral types, $\mathcal{S}_{3}$ and $\mathcal{S}_{4}$ in that order, that appear in region RII (solid black lines), and some of the center type, $\mathcal{S}_{1}$ and $\mathcal{S}_{2}$, which disappear in RV (dashed blue lines).} 
	\label{fig:Fig3}
\end{figure}

Now, we add linear gain or loss to the previous model, making $\alpha_{j}$ complex numbers, hence the dynamical system becomes,
\begin{eqnarray}
\partial_{z} S_{x} &=& \left( 2 n_{i} S_{x} - a S_{y} \right) S_{z}  , \nonumber \\ 
\partial_{z} S_{y} &=& \left( 2 g + a S_{x}  + 2 n_{i} S_{y} \right) S_{z}, \nonumber \\
\partial_{z} S_{z} &=& -2 g S_{y} - 2 n_{i} \left( 1 - S_{z}^2 \right),  \label{eq:WeakFiberAmplifier}
\end{eqnarray}
and gives rise to the following fixed points,
\begin{eqnarray}
\mathcal{S}_{1,2} &=& \left( \pm \sqrt{1- \left(\frac{n_{i}}{g}\right)^2}, -\frac{n_{i}}{g},0\right), \nonumber \\
\mathcal{S}_{3,4} &=&\left( - \frac{a f^2 }{2 g}, -\frac{n_{i} f^2}{g},\pm \sqrt{ 1 - f^2} \right), 
\end{eqnarray}
where  we have defined $f^2 =  4 g^2 / \left( 4 n_{i}^{2} + a^{2} \right)$.
The Jacobian eigenvalues evaluated at each of these fixed points are too cumbersome to write them here explicitly, but Table \ref{tab:Tab2} presents their classification. 
The inclusion of gain or loss modifies the dynamics considerably. 
We now find two centers, $\mathcal{S}_{1}$ and $\mathcal{S}_{2}$, in a region that we call RI, defined by the conditions $n_{i} < g$ and $d^2 > 1$.
In another (codimension-one) region, RII where $n_{i} < g$ and $d^2 = 1$, the fixed point $\mathcal{S}_{1}$ is still a center, and we have a triple-degenerate nongeneric fixed point, $\mathcal{S}_{2}=\mathcal{S}_{3}=\mathcal{S}_{4}$.
Here, point $S_{2}$ turns from a center in region RI into a saddle point in region RIII, which is defined by parameters $n_{i} < g$ and $d^2 < 1$, similar to what is shown above for the passive system.
The two other points, $\mathcal{S}_{3}$ and $\mathcal{S}_{4}$, transform into unstable and stable spiral points, respectively, in region RIII. 
We further define another region of codimension one as RIV with $n_{i} = g$.
Here, the two initial centers $\mathcal{S}_{1}$ and $\mathcal{S}_{2}$ coalesce into a single fixed point that disappears in region RV, defined by $n_{i} > g$.
Figure \ref{fig:Fig3}(a) shows these five regions in the parameter space, and Fig.~\ref{fig:Fig3}(b) presents the appearance of the spiral fixed points, $\mathcal{S}_{3}$ and $\mathcal{S}_{4}$, see solid black lines, and the disappearance of the centers, $\mathcal{S}_{1}$ and $\mathcal{S}_{2}$, see dashed blue lines. 
\begin{table}[htbp]
	\centering
	\caption{\bf Fixed Point Classification for the Twisted Weakly Birefringent Fiber Amplifier Model Based on Eq.(\ref{eq:WeakFiberAmplifier})}
	\begin{tabular}{cccccc}
		\hline
		& RI & RII & RIII & RIV & RV  \\
		& $  n_{i} < g $ & $  n_{i} < g $ & $  n_{i} < g $ &  $n_{i}=g$  & $  n_{i} > g $ \\
		& $f^2 > 1$ & $f^2 = 1$ & $f^2 < 1$  \\
		\hline
		$\mathcal{S}_{1}$ & Center  & Center 		& Center 	& Nongeneric &  \\
		$\mathcal{S}_{2}$ & Center  & Nongeneric	& Saddle 	& Nongeneric &  \\
		$\mathcal{S}_{3}$ &   		& Nongeneric	& Repellor	& Repellor	& Repellor \\
		$\mathcal{S}_{4}$ &   		& Nongeneric	& Node 		& Node		& Node\\
		\hline
	\end{tabular}
	\label{tab:Tab2}
\end{table}

\begin{figure}[htbp]
	\centering
	\fbox{\includegraphics{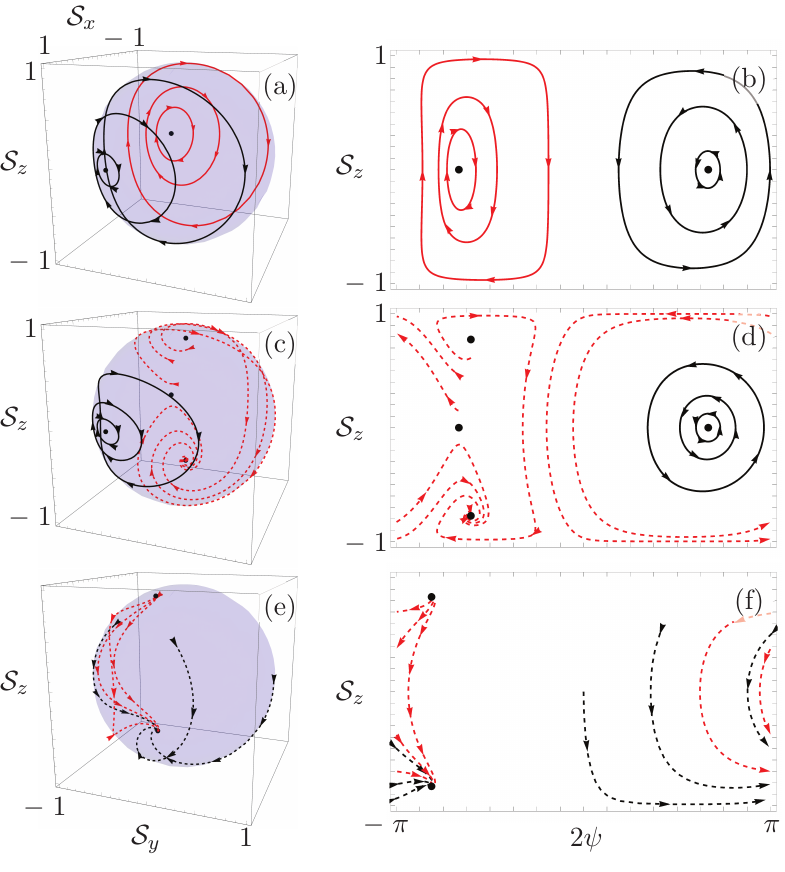}}
	\caption{(Color online) Polarization trajectories on (a),(c),(e) the Poincar\'e sphere and (b),(d),(f) its cylindrical projection, for a weakly birefringent twisted nonlinear fiber, (a) and (b) RI for $n_{i} = 0.5 g$ and  $a=g$,  (c) and (d) RIII for $n_{i} = 0.5 g$ and $a=3g$, (e) and (f) RV for $n_{i} = 1.5 g$ and $a = 2 g$. Rabi oscillations are shown by solid lines and dashed lines represent trajectories attracted to the spiral node. } 
	\label{fig:Fig4}
\end{figure}

Now, the polarization circulates around two centers, $\mathcal{S}_{1}$ and $\mathcal{S}_{2}$, in RI, as shown by solid lines in Figs. \ref{fig:Fig4}(a) and \ref{fig:Fig4}(b).
The center $\mathcal{S}_{1}$ persists in RI through RIII, as shown by solid black lines in Figs. \ref{fig:Fig4}(c) and \ref{fig:Fig4}(d), but the fixed point $\mathcal{S}_{2}$ turns into a saddle point in region RIII, with the difference from the passive model being the lack of Rabi oscillations due to the unstable and stable spiral points, $\mathcal{S}_{3}$ and $\mathcal{S}_{4}$, that repel and attract, respectively, polarization trajectories on the PS.
In region RIII, initial states with polarization-ellipse angle values in the range $0 < 2 \psi < \pi/2$ initiate polarization circulation along stable elliptical orbits on the PS, that maintain the polarization angle in the same range, the solid black lines in Figs. \ref{fig:Fig4}(c) and \ref{fig:Fig4}(d).
This is not true for initial states with polarization angle values in the range 
$-\pi/2 < 2 \psi < 0$ where polarization trajectories may run over a larger portion of the PS, avoiding the unstable spiral point $\mathcal{S}_{3}$, before being trapped by the stable spiral point $\mathcal{S}_{4}$, dashed red lines in Figs. \ref{fig:Fig4}(c) and \ref{fig:Fig4}(d).
Once the centers coalesce, as they reach the boundary (codimension-one) of RIV, and disappear in region RV, the dynamics become simpler, with initial polarization states following trajectories on the PS that avoid the unstable spiral fixed point $\mathcal{S}_{3}$ and are eventually trapped by the attractor $\mathcal{S}_{4}$, dashed lines in Figs. \ref{fig:Fig4}(e) and \ref{fig:Fig4}(f).
The latter regime can be used to design a fixed-polarization amplifier.

\section{Conclusions}
In conclusion, we have elaborated a model for a twisted birefringent fiber amplifier with saturable nonlinearity in the continuous-wave regime. Our model may serve as an optical simulation of the semi-classical non-Hermitian Bose-Hubbard dimer, known in condensed matter physics. We showed that, assuming weak birefringence and a particular twisting rate, our model is tantamount to the one for twisted passive fibers, and that it provides novel dynamical behavior when gain or loss is included. While our examples refer to realizations that allow us to construct analytic closed-form solutions, our approach applies equally well to any given parameter set that an experimental realization may provide.

Our fiber amplifier model may help in the design of polarization circulators that maintain the polarization-ellipse angle in the initial state range, polarization filters that differentiate between different values of the initial polarization angle, and polarization amplifiers that are insensitive to the initial polarization, to mention a few practical examples. The realization of these applications is controlled by the interplay between the birefringence, mechanical twist, effective gain, mode coupling and Kerr nonlinearity.
Future work might include four-wave mixing, which is a well-known platform for polarization control in optical fiber systems   \cite{Assemat2010p2025,Fatome2012p938,Guasoni2012p1511,Guasoni2012p2710,Kozlov2013p530}, in addition to the saturated self-phase modulation considered here. \\\\
\hspace{0em}
\textbf{Funding} Consejo Nacional de Ciencia y Tecnolog\'ia (CONACYT) (294921, FORDECYT 290259); Israel Science Foundation (ISF) (1286/17). 

\vspace{6pt} 


\ack{J. D. Huerta Morales acknowledges funding from CONACYT. B. M. Rodr\'iguez-Lara thanks E. Kuzin for fruitful discussion and acknowledges support from Photonics and Mathematical Optics Group at Tecnol\'ogico de Monterrey and Consorcio en \'Optica Aplicada through CONACYT. B. A. Malomed appreciates hospitality of Escuela de Ingenier\'ia y Ciencias at Tecnol\'ogico de Monterrey.  }

\section*{References}
\bibliographystyle{unsrt}

\end{document}